# Low-Frequency 1/*f* Noise in MoS$_2$ Transistors

J. Renteria[1,#], R. Samnakay[2,#] S. L. Rumyantsev[3,4], P. Goli[1,2], M.S. Shur[3] and A.A. Balandin[1,2,*]

[1]Nano-Device Laboratory, Department of Electrical Engineering, Bourns College of Engineering, University of California – Riverside, Riverside, California 92521 USA

[2]Materials Science and Engineering Program, Bourns College of Engineering, University of California – Riverside, Riverside, California 92521 USA

[3]Department of Electrical, Computer, and Systems Engineering, Center for Integrated Electronics, Rensselaer Polytechnic Institute, Troy, New York 12180, USA

[4]Ioffe Physical-Technical Institute, St. Petersburg 194021, Russia

**Abstract**

We report on the results of the low-frequency (1/*f*, where *f* is frequency) noise measurements in MoS$_2$ field-effect transistors revealing the relative contributions of the MoS$_2$ channel and Ti/Au contacts to the overall noise level. The investigation of the 1/*f* noise was performed for both as fabricated and aged transistors. It was established that the McWhorter model of the carrier number fluctuations describes well the 1/*f* noise in MoS$_2$ transistors, in contrast to what is observed in graphene devices. The trap densities extracted from the 1/*f* noise data for MoS$_2$ transistors, are $1.5 \times 10^{19}$ eV$^{-1}$cm$^{-3}$ and $2 \times 10^{20}$ eV$^{-1}$cm$^{-3}$ for the as fabricated and aged devices, respectively. It was found that the increase in the noise level of the aged MoS$_2$ transistors is due to the channel rather than the contact degradation. The obtained results are important for the proposed electronic applications of MoS$_2$ and other van der Waals materials.

#These authors contributed equally to research.

*Corresponding author (AAB): balandin@ee.ucr.edu

Recent advances in the exfoliation and growth of two-dimensional (2D) layered materials have allowed for investigation of their electronic and optical properties [1-4]. Among these material systems, molybdenum disulfide ($MoS_2$) is one of the more stable layered transition-metal dichalcogenides (TMDCs) [5-6]. Each layer of $MoS_2$ consists of one sub-layer of molybdenum sandwiched between two other sub-layers of sulfur in a trigonal prismatic arrangement [7]. A single-layer $MoS_2$ shows a direct band gap of 1.8 eV, while bi-layer and bulk $MoS_2$ exhibit an indirect band gap of 1.3 eV and 1.2 eV, respectively [8-9]. It has been demonstrated that bi- and few-layer $MoS_2$ devices are promising for sensing, optoelectronic, and energy harvesting applications [10-12]. Owing to its relatively large energy band gap, the $MoS_2$ field-effect transistors (FETs) offer reasonable on-off ratios, which suggests possibilities for digital or analog electronic applications of this 2D *van der Waals* material [12-13].

Like with any other material system, practical applications of $MoS_2$ devices in sensing and in digital or analog electronics are only possible if the material and devices meet the minimum level requirements for the low-frequency $1/f$ noise [14-22]. The sensitivity of amplifiers and transducers used in sensors is ultimately defined by the flicker ($1/f$) noise [22]. The accuracy of a system limited by $1/f$ noise cannot be improved by extending the measuring time, $t$, because the total accumulated energy of the $1/f$ noise increases at least as fast as the measuring time $t$. In contrast, the system accuracy limited by a white noise, e.g. shot or thermal noise, increases the measuring time as $t^{1/2}$. For this reason, the sensitivity and selectivity of many types of sensors, particularly those that rely on electrical response, is limited by $1/f$ noise. Although $1/f$ noise dominates the noise spectrum only at low frequencies, its level is equally important for electronic applications at high frequencies, because $1/f$ noise is the major contributor to the phase noise of the oscillating systems. The up-conversion of $1/f$ noise is a result of unavoidable non-linearity in devices and the electronic systems, which leads to phase noise contributions.

Meeting the requirements for $1/f$ noise level could be particularly challenging for 2D materials, where the electrons in the conducting channels are ultimately exposed to the charged traps in the gate dielectrics and substrates [23]. The contributions of contacts to the low-frequency noise can also be significant owing to imperfection of the technology for metal deposition on TMDCs. Investigations of the low-frequency $1/f$ noise in $MoS_2$ devices are in its infancy [24-25], and many questions regarding the specific physical mechanism of 1/*f* noise in this material remained unanswered, including the role of metal contacts and aging. The nanometer-scale thickness of the device channel may change the noise level compared to devices with conventional feature sizes [18-23]. In this letter, we address these issues while focusing on separating the contributions from the $MoS_2$ channel and Ti/Au contacts to the overall noise level. The devices selected for this study used bi-layer and tri-layer $MoS_2$ films, because they are more robust for practical electronic applications.

Thin films of $MoS_2$ were exfoliated from bulk crystals and transferred onto $Si/SiO_2$ substrates following the standard "graphene-like" approach [26-28]. The thickness $H$ of the films ranged from bi-layer to a few tri-layers. Micro-Raman spectroscopy (Renishaw InVia) verified the crystallinity and thickness of the flakes after exfoliation. It was performed in the backscattering configuration under $\lambda$=488-nm laser excitation laser using an optical microscope (Leica) with a 50× objective. The excitation laser power was limited to less than 0.5 mW to avoid local heating. In Figure 1, we present informative bands at ~383.2 cm$^{-1}$ ($E^1_{2g}$) and 406.5 cm$^{-1}$ ($A_{1g}$), consistent with the previous reports of the $MoS_2$ Raman spectrum [29]. Analysis of the Raman spectrum indicates that this sample is a tri-layer $MoS_2$ film. The latter follows from the frequency difference, $\Delta\omega$, between the $E^1_{2g}$ and the $A_{1g}$ peaks. The increase in the number of layers in $MoS_2$ films is accompanied by the red shift of the $E^1_{2g}$ and blue shift of the $A_{1g}$ peaks [29]. This sensitivity of the Raman spectral features of $MoS_2$ to the film thickness was used to reliably determine the thickness of the samples used for fabricating FETs.

[Figure 1: Raman]

Devices with $MoS_2$ channels were fabricated using electron beam lithography (LEO SUPRA 55) for patterning of the source and drain electrodes and the electron-beam evaporation (Temescal BJD-1800) for metal deposition. Conventional Si substrates with 300-nm thick $SiO_2$ layers were spin coated (Headway SCE) and baked consecutively with two positive resists: first, methyl methacrylate (MMA) and then, polymethyl methacrylate (PMMA). These devices consisted of $MoS_2$ thin-film channels with Ti/Au (10-nm / 100-nm) contacts. The heavily doped $Si/SiO_2$ wafer served as a back gate. Figure 2 (a-b) shows optical microscopy and scanning electron microscopy (SEM) images of representative MoS$_2$ – Ti/Au devices. The majority of the bi-layer and tri-layer thickness devices had a channel length, $L$, in the range from 1.3 µm to 3.5 µm, and the channel width, $W$, in the range from 1 µm to 7 µm.

[Figure 2: Devices]

Figure 3 (a-c) shows the room-temperature (RT) current-voltage (I-V) characteristics of the fabricated MoS$_2$ devices.  Figure 3 (a) presents repeated sweeps of the source-drain voltage in the range from -0.1 V to +0.1 V. The linear I-V characteristics suggest that the MoS$_2$ – Ti/Au contacts are Ohmic. Figures 3 (b) and (c) show the drain-source current, $I_{ds}$, as a function of the back-gate bias, $V_g$, in the semi-log and linear scale, respectively. As seen, the device behaves as an $n$-channel field effect transistor. The curves of different colors correspond to the source-drain bias, $V_{ds}$, varying from 10 mV to 100 mV. As seen from Figure 3 (b), a representative device reproducibly reveals a well-defined threshold voltage, $V_{th}$=(-7) – (-8) V obtained from the linear extrapolation of $I_d$ versus $V_g$ characteristics (in the linear scale). The threshold voltage varied from device to device depending on channel size. It steadily shifted more negative as a result of aging. The current on/off ratio greater than $6.6 \times 10^3$ was determined at a drain-source bias of 80 mV. We deduced a subthreshold slope of 549 mVdec$^{-1}$ at the bias of $V_{ds}$ = 100 mV. The mobility values for these devices were in the range of 1 - 4 cm$^2$/Vs, which are typical for similarly fabricated MoS$_2$ FETs [8-11]. The low-field mobility was extracted using the formula [30] $\mu = (dI_{ds}/dV_g) \cdot (L/(WC_{ox}V_{ds}))$, where $C_{OX} = \varepsilon_o \varepsilon_r / d$ = 1.15 $\times 10^{-4}$ (F/m$^2$) is the oxide capacitance (where $\varepsilon_o$ is the dielectric permittivity of free space, $\varepsilon_r$ is the relative dielectric permittivity and $d$ is the oxide thickness). We used $\varepsilon_r$=3.9 and $d$=300 nm for SiO$_2$. As seen from Figure 3 (c) the current voltage characteristics demonstrate a knee at $V_g$~-2.5 V but do not tend to saturate at least up to $V_g$=-20V. The minimum drain to source resistance at $V_g$=20 V is ~2.5 MΩ Therefore the contact resistance should be significantly smaller.

[Figure 3: I-Vs]

The noise was measured in the linear region at $V_d$=50 mV keeping the source at the ground potential. The voltage fluctuations from the drain load resistance of $R_L$=50 kΩ were analyzed with a dynamic signal analyzer (SR785). The measurements were conducted under ambient conditions at room temperature. Figure 4 shows typical low-frequency noise spectra of voltage fluctuations, $S_v$, as a function of frequency for several values of drain-source and gate biases. One can see that the low-frequency noise is of the 1/$f$ type without any signatures of generation-recombination bulges. To verify how closely the noise spectral density follows 1/f dependence, we fitted the experimental data with 1/$f^\alpha$. The parameter $\alpha$ varied in the range from ~0.75 to ~1.25 without revealing any clear gate bias, $V_g$, dependence. The latter suggests that the traps contributing to the noise distributed uniformly in space and energy [22].

[Figure 4 : Voltage Noise]

For any new material technology it is important to analyze the relative contributions of the device channel and contacts as well as to assess the effects of aging. To accomplish this goal, we calculated the short-circuit current fluctuations in the usual way as $S_I=S_v[(R_L+R_D)/(R_L R_D)]^2$, where $R_L$ and $R_D$ are the load and device resistances, respectively. The noise spectrum density at different drain-source biases was consistently proportional to the current squared at a constant gate voltage $V_g$: $S_I \sim I_{ds}^2$. The latter implies that the current does not drive the fluctuations but merely makes the fluctuations in the sample visible via Ohm's law [18]. The noise was measured in the same devices in the span of two weeks. As a result of aging the threshold voltage shifted ~1 V to more negative value and total drain to source resistance increased ~20%. The circular symbols in Figure 5 represent the normalized current noise, $S_I/I_{ds}^2$, as a function of the gate bias for the as fabricated device (black symbols) and device aged for a week in ambient atmosphere (red symbols). One can see that the normalized noise spectral density is an order of magnitude larger in the week old device. The latter suggests that capping of $MoS_2$ with some protective layer may be a technologically viable way for reducing 1/$f$ noise for practical applications.

[Figure 5: Current Noise vs Gate]

Let us now investigate the relative contribution of the metal contacts and device channels to the overall level of 1/$f$ noise. This issue is of particular importance for MoS$_2$ devices due to the fact that the technology of metal contact fabrication to TMDCs is still rudimentary. Since the contact resistance is not negligible we consider that both the metal contact and MoS$_2$ channel contribute to the measured noise. In this case, we can write that [22]

$$\frac{S_I}{I^2} = \frac{S_{RCh}}{R_{CH}^2} \frac{R_{Ch}^2}{(R_{Ch}+R_C)^2} + \frac{S_{RC}}{R_C^2} \frac{R_C^2}{(R_{Ch}+R_C)^2}. \qquad (1)$$

Here $S_{RCh}/R^2_{Ch}$ is the noise spectral density of the channel resistance fluctuations, $R_{Ch}$ is the resistance of the channel, $R_C$ is the contact resistance, and $S_{Rc}/R_c^2$ is the noise spectral density of the contacts resistance fluctuations.

Assuming that the channel noise complies with the McWhorter carrier number fluctuation model, we can write for the noise spectral density, $S_{RCh}/R^2_{Ch}$, the following equation [14]

$$\frac{S_{RCh}}{R_{CH}^2} = \frac{kTN_t}{\gamma f W L n_s^2}, \qquad (2)$$

where $k$ as the Boltzmann constant, $T$ is the temperature, $\gamma$ is the tunneling parameter taken to be $\gamma=10^8$ cm$^{-1}$, $n_s$ is the channel concentration, and $N_t$ is the trap density. In the strong inversion regime, the concentration $n_s$ can be estimated as $n_s=C_{ox}(V_g-V_t)/q$. Here $C_{ox}$ is the gate capacitance per unit area, and $q$ is the elemental charge. Since the total resistance, $R_{ch}+R_c$, was measured directly, there are three fitting parameters in our analysis, $N_t$, $S_{Rc}/R_c^2$, and $R_C$. In Figure 5, we show with the dash lines the model fitting for the noise dominated by the channel contribution and, separately, by the contact contribution (i.e. the first term and the last term in Eq. (1), respectively). The solid lines show the sum of both contributions. The fitted values of the contact noise and contact resistance are determined to be $S_{Rc}/R_c^2=10^{-4}/f$, $R_C=10^6\,\Omega$ for both as fabricated and aged samples. The extracted trap densities are $N_t=1.5\times10^{19}$ eV$^{-1}$cm$^{-3}$ and $N_t=2\times10^{20}$ eV$^{-1}$cm$^{-3}$ for as fabricated and aged samples, respectively.

The excellent agreement of the model fitting (Eq. (1) and Eq. (2)) with the experimental results indicates that the *a priori* assumption of the McWhorter model description was valid. The model description allows one to clearly distinguish the contributions to the noise from the MoS$_2$ channel and from the metal contacts. The absolute value of the trap density extracted is within the range typical for conventional FETs. For example, the trap densities determined for Si metal-oxide-semiconductor field-effect transistors (MOSFETs) with high-k dielectric were $N_t=10^{18}$ - $10^{20}$ (eV$^{-1}$cm$^{-3}$) [31-33] while those for GaN-based heterostructures field-effect transistors (HFETs) were $N_t=10^{18}$ - $4\times10^{20}$ (eV$^{-1}$cm$^{-3}$) [34].

Let us now compare the noise mechanism in MoS$_2$ thin-films with that in conventional semiconductors, metals and graphene devices. It is known that 1/*f* noise is either due to the mobility fluctuations or the number of carriers fluctuations. In conventional semiconductor devices, such as Si complementary metal-oxide-semiconductor (CMOS) field-effect transistors (FETs), 1/*f* noise is described by the McWhorter model [14], which is based on the carrier-number fluctuations. In metals, on the other hand, 1/*f* noise is usually attributed to the mobility fluctuations [18]. There are materials and devices where contributions from both mechanisms are comparable or cross-correlated. By assuming the McWhorter model for the MoS$_2$ channel noise we were able to successfully reproduce the overall noise gate-bias dependence in MoS$_2$ FETs. The latter indicates that the 1/*f* noise mechanism in MoS$_2$ FETs is similar to that in conventional Si CMOS transistors: carrier number fluctuations with the traps widely distributed in space and energy. It is important to note here such 1/*f* noise behavior is quite different from that of another important 2D material, graphene, where the gate voltage dependence of noise does not follow the McWhorter model [14]. It was shown that 1/*f* noise in graphene can be more readily described by the mobility fluctuation [23]. The latter was concluded on the basis of analysis of the gate bias dependence [35-37], effect of electron beam irradiation damage [38], noise scaling with the thickness [39-40] and measurements of noise in graphene devices under magnetic field [41].

There is another important observation from experimental data and model fitting presented in Figure 5. Since the experimental data for as fabricated and aged devices is fitted well with the same value of the contact noise and contact resistance we can conclude that aging results mainly from the deterioration of the MoS$_2$ channel. The degradation in the channel property is expressed via the value of the effective trap density, $N_t$. The latter increased by more than an order of magnitude after environmental exposure for one week. The fitting of the noise data provides an indirect estimate for the contact resistance. We also note that since the noise in the tested FETs with nanometer-scale $MoS_2$ channels is described by the carrier number fluctuation model, the use of the Hooge parameter, introduced specifically for the mobility fluctuation noise in bulk conductors, is not well justified from the physics point of view.

In conclusion, we reported results of the low-frequency noise investigation in $MoS_2$ FETs with Ti/Au contacts. It was established that both the channel and contacts contribute to the overall $1/f$ noise level of the as fabricated and aged transistors. The intrinsic noise characteristics in $MoS_2$ devices are well described by the McWhorter model of the carrier number fluctuations, in contrast to graphene devices. It was found that the increase in the noise level in aged $MoS_2$ transistors is due to channel rather than contact degradation. The obtained results can be used for optimization of the devices with the channels implemented with $MoS_2$ and other van der Waals materials.

*Acknowledgements*

This work was supported, in part, by the Semiconductor Research Corporation (SRC) and Defense Advanced Research Project Agency (DARPA) through STARnet Center for Function Accelerated nanoMaterial Engineering (FAME). AAB and RKL also acknowledge funding from the National Science Foundation (NSF) and SRC Nanoelectronic Research Initiative (NRI) for the project 2204.001: Charge-Density-Wave Computational Fabric: New State Variables and Alternative Material Implementation (NSF ECCS-1124733) as a part of the Nanoelectronics for 2020 and beyond (NEB-2020) program. AAB also acknowledges funding from NSF for the project Graphene Circuits for Analog, Mixed-Signal, and RF Applications (NSF CCF-1217382). SLR acknowledges partial support from the Russian Fund for Basic Research (RFBR). The work at RPI was supported by the National Science Foundation under the auspices of the EAGER program.

## FIGURE CAPTIONS

**Figure 1:** Raman spectrum of a $MoS_2$ thin film showing the $E^1_{2g}$ and the $A_{1g}$ peaks. The increase in the number of layers in $MoS_2$ films is accompanied by the red shift of the $E^1_{2g}$ and blue shift of the $A_{1g}$ peaks. The energy difference, $\Delta\omega$, between $E^1_{2g}$ and the $A_{1g}$ peaks indicates that the given sample is a tri-layer $MoS_2$ film.

**Figure 2:** Scanning electron microscopy image of a representative $MoS_2$ – Ti/Au field-effect transistor. The pseudo colors are used for clarity: yellow corresponds to the metal contacts while blue corresponds to $MoS_2$ thin-film channel.

**Figure 3:** Current-voltage characteristics of the fabricated $MoS_2$ FET at room temperature. The drain-source current for repeated sweeps of the source-drain voltage in the small-voltage range between -0.1 V and +0.1 V at $V_g$=0 V (a). The drain-source current, $I_{ds}$, shown as a function of the back-gate bias, $V_g$, in the semi-log (b), and linear scale (c). The inset shows an optical image of the measured device.

**Figure 4:** Typical low-frequency noise spectra of voltage fluctuations, $S_v$, as a function of frequency f for different values of the gate bias. The data are for the linear regime at $V_d$=50 mV and the source contact at a ground potential. The measurements were conducted under ambient conditions at room temperature.

**Figure 5:** Measured and simulated low-frequency noise response of $MoS_2$ FETs. The circular symbols represent the experimental data points for the normalized current noise spectral density, $S_I/I_{ds}^2$, as a function of the gate bias for the as fabricated device (black symbols) and device aged for a week under ambient conditions (red symbols). The normalized noise spectral density is an order of magnitude larger in the week old device. The dashed lines represent the model fitting for the noise dominated by the channel contribution and, separately, by the contact contribution. The solid lines show the sum of both contributions. The agreement between the theoretical fitting and experimental data indicate that the 1/*f* noise in $MoS_2$ FETs follow the carrier number fluctuation model.

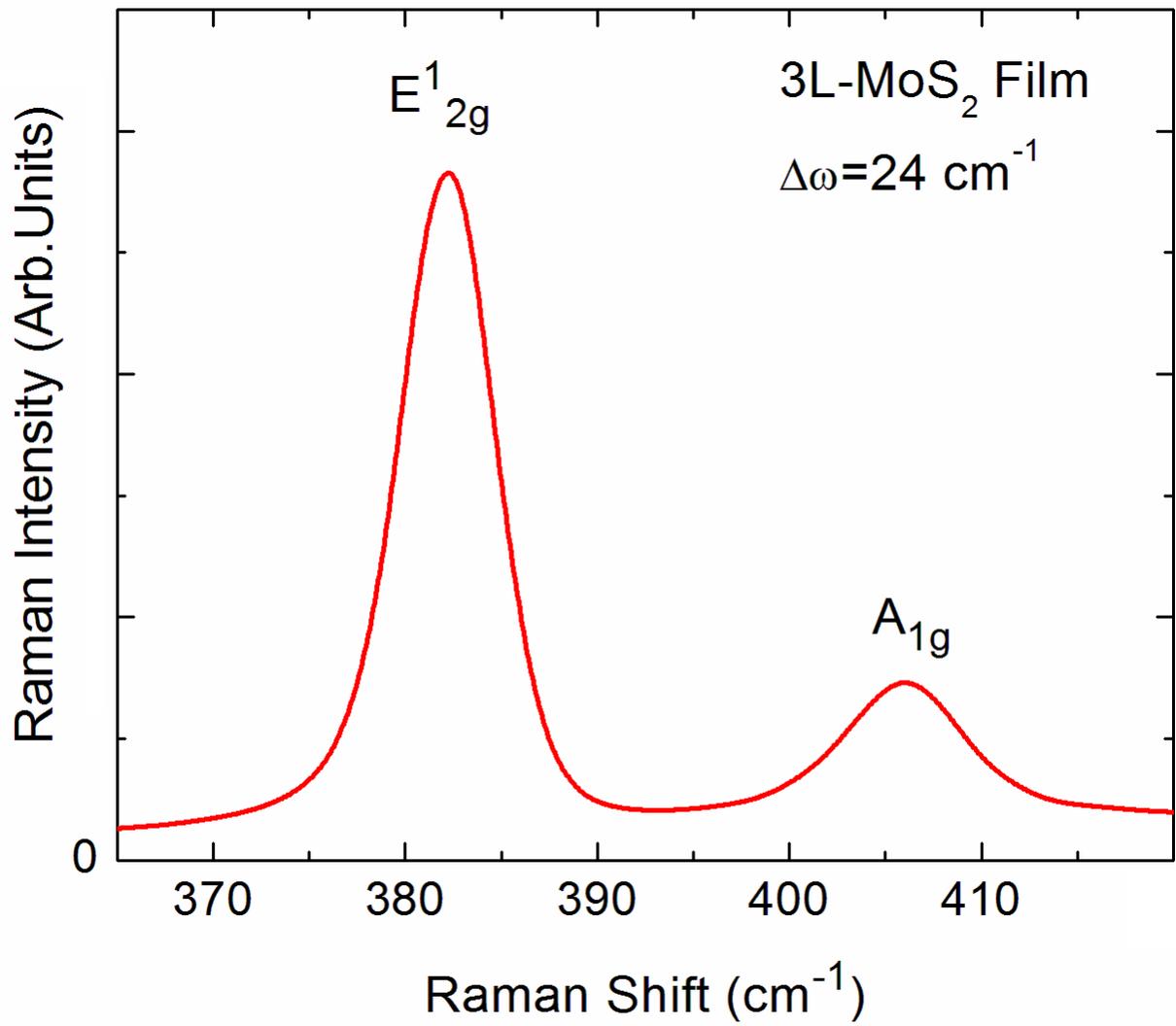

Figure 1

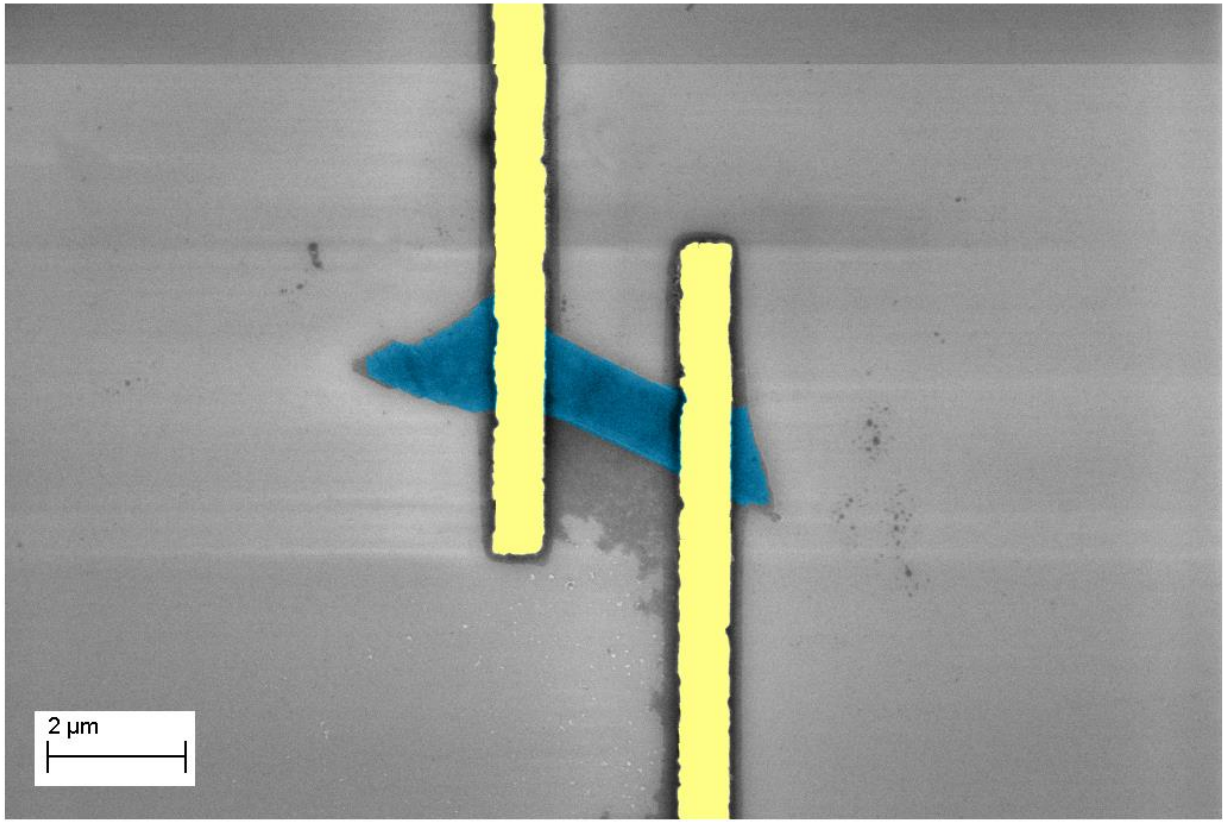

Figure 2

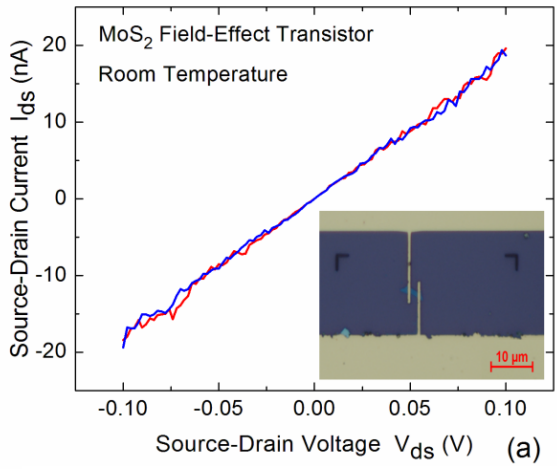

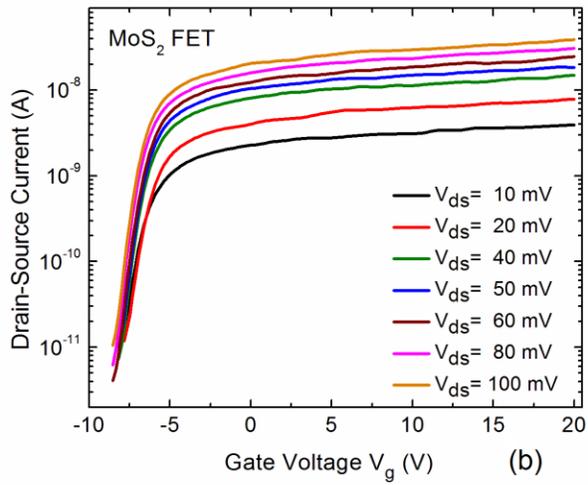

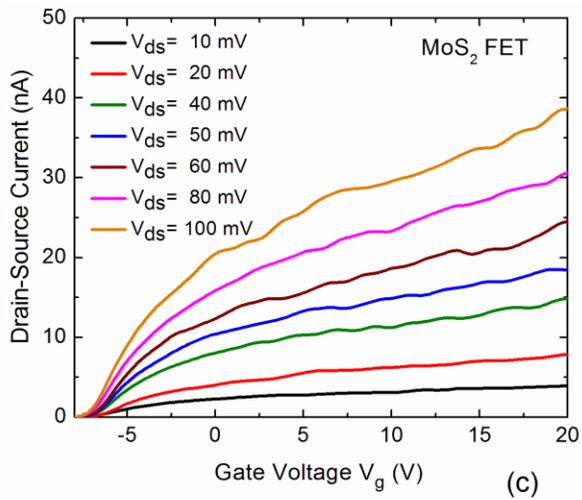

Figure 3

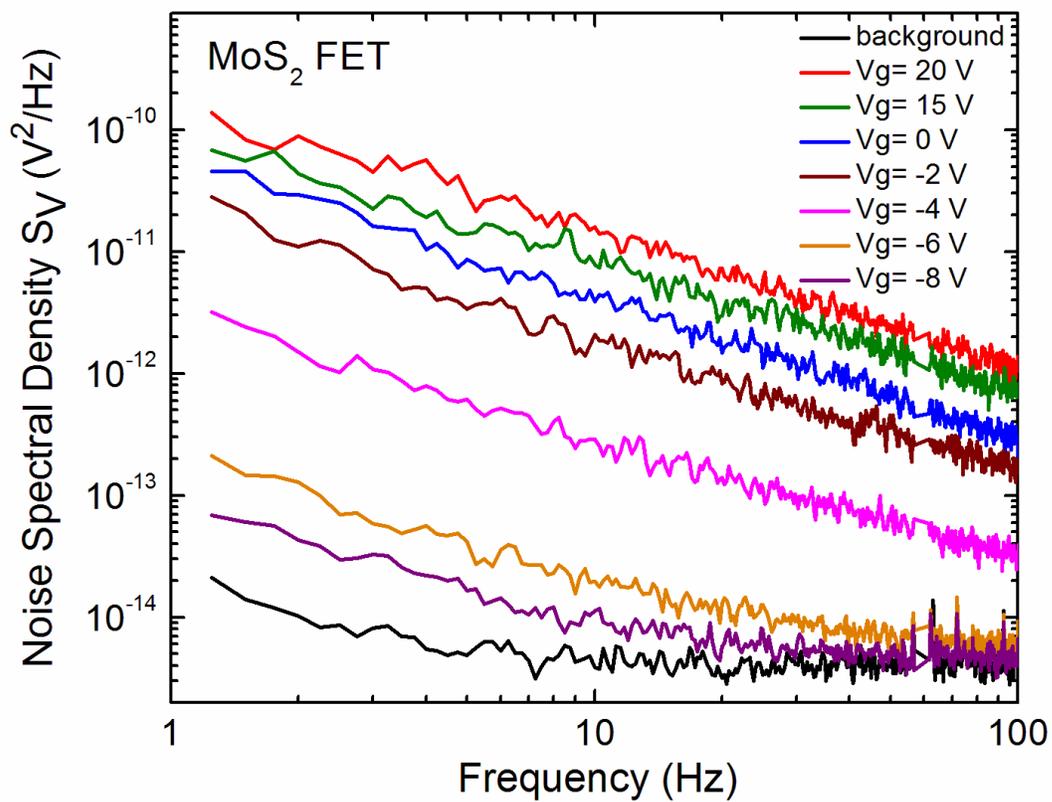

Figure 4

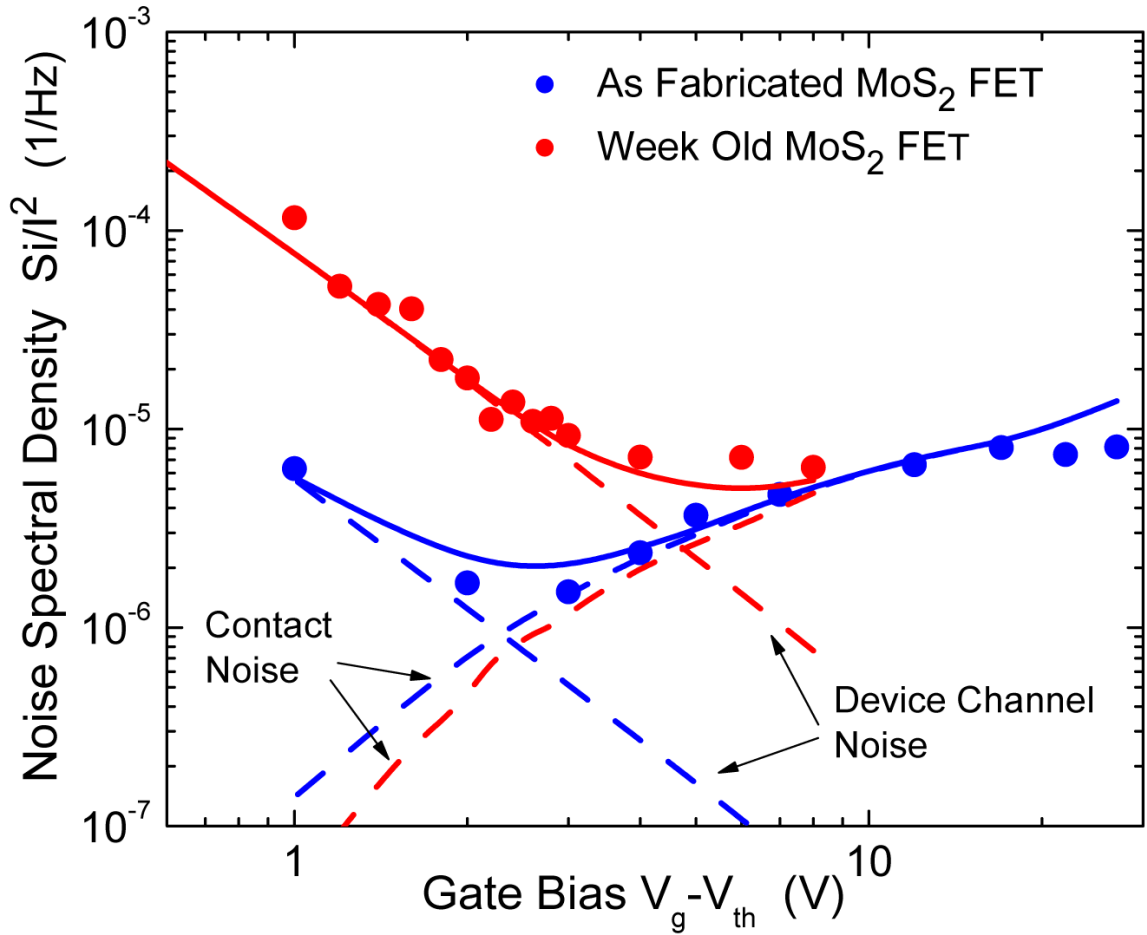

Figure 5